\documentclass{iitpressproc}

\usepackage{graphicx}
\usepackage{amsmath}

\def\bea{\begin{eqnarray}}
\def\eea{\end{eqnarray}}

\def\pp{\mbox{$p$-$p$}}
\def\ee{\mbox{$e^+$-$e^-$}}

\def\auau{\mbox{Au-Au}}

\def\aa{\mbox{A-A}}

\begin{document}

\title{Improved isolation of the p-p underlying event based on minimum-bias trigger-associated hadron correlations}

\author{Thomas A.\ Trainor and Duncan J.\ Prindle
\address{CENPA 354290, University of Washington, Seattle, USA}\\[2ex]
}

\maketitle

\begin{abstract}
Some aspects of hadron production in \pp\ collisions remain unresolved, including the low-hadron-momentum structure of high-parton-energy dijets, separation of triggered dijets from the {\em underlying event} (UE), the systematics of {multiple parton interactions} and possible systematic underestimation of dijet contributions to high-energy nuclear collisions.
In this study we apply a minimum-bias {\em trigger-associated} (TA) correlation analysis to \pp\ collisions. We extract a hard component from TA correlations that can be compared with measured jet fragment systematics derived from \ee\ collisions. The kinematic limits on jet fragment production may be determined.
 The same method may be extended to \aa\ collisions where the role of minimum-bias jets in spectra and correlations is strongly contested.
\end{abstract}

\section{Introduction} 

Several open issues for hadron production in \pp\ collisions relate to dijet production, both the frequency of hard parton scattering and the subsequent fragmentation to jets. In this study we infer the hard scattering rate from the two-component multiplicity systematics of single-particle spectra and introduce a trigger-associated correlation analysis to extract minimum-bias jet fragment distributions. We wish to determine the  momentum correlation structure of minimum-bias jets down to the kinematic limits.

\section{Two-component model of p-p single-particle $\bf y_t$ spectra}

The two-component model of single-particle (SP) spectra is defined by~\cite{ppprd}
\bea \label{ppspec}
{dn_{ch}}/{y_t dy_t \Delta \eta} 
&=&  \rho_s( n_{ch}) S_0(y_t)  +  \rho_{h}( n_{ch}) H_0(y_t),~~~
\eea
where $n_{ch}$ is integrated within some acceptance $\Delta \eta$ and $\rho_x = n_x / \Delta \eta$. 
\begin{figure}[h]
\centering
\includegraphics[width=0.24\columnwidth]{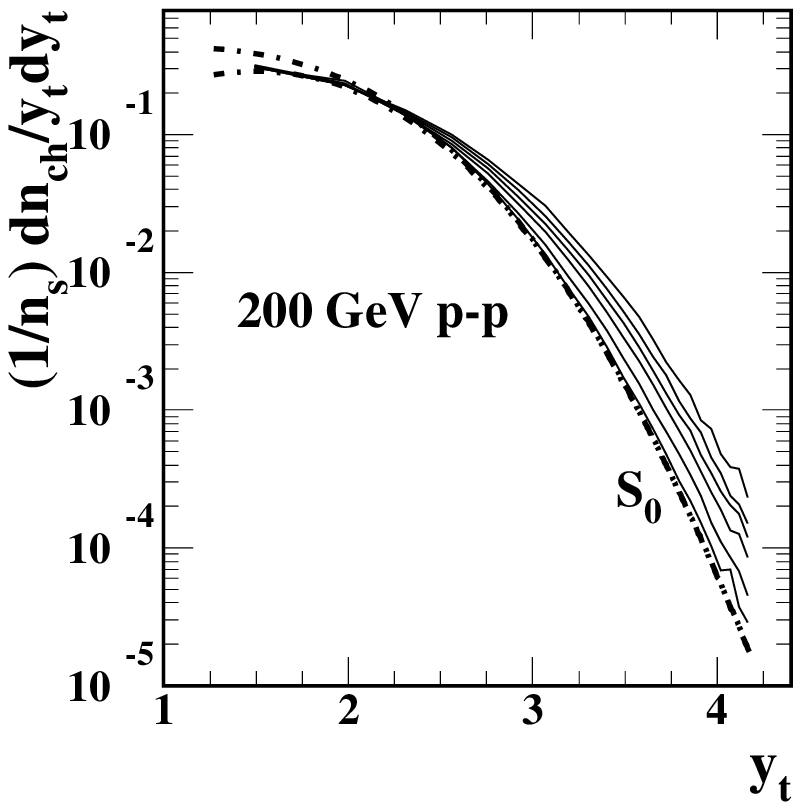}
\includegraphics[width=0.24\columnwidth]{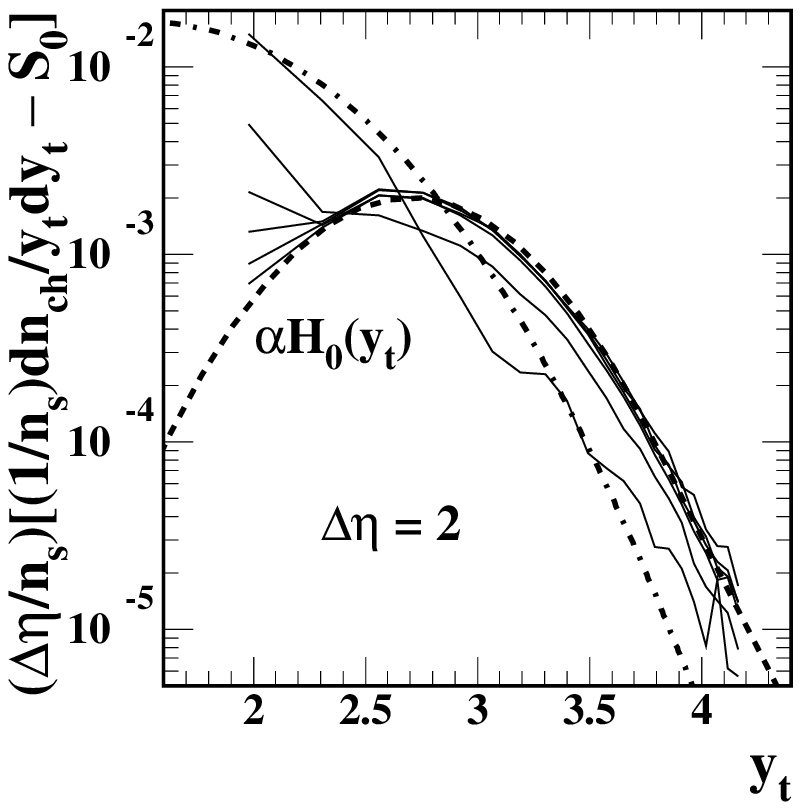}
\includegraphics[width=0.24\columnwidth,height=0.242\columnwidth]{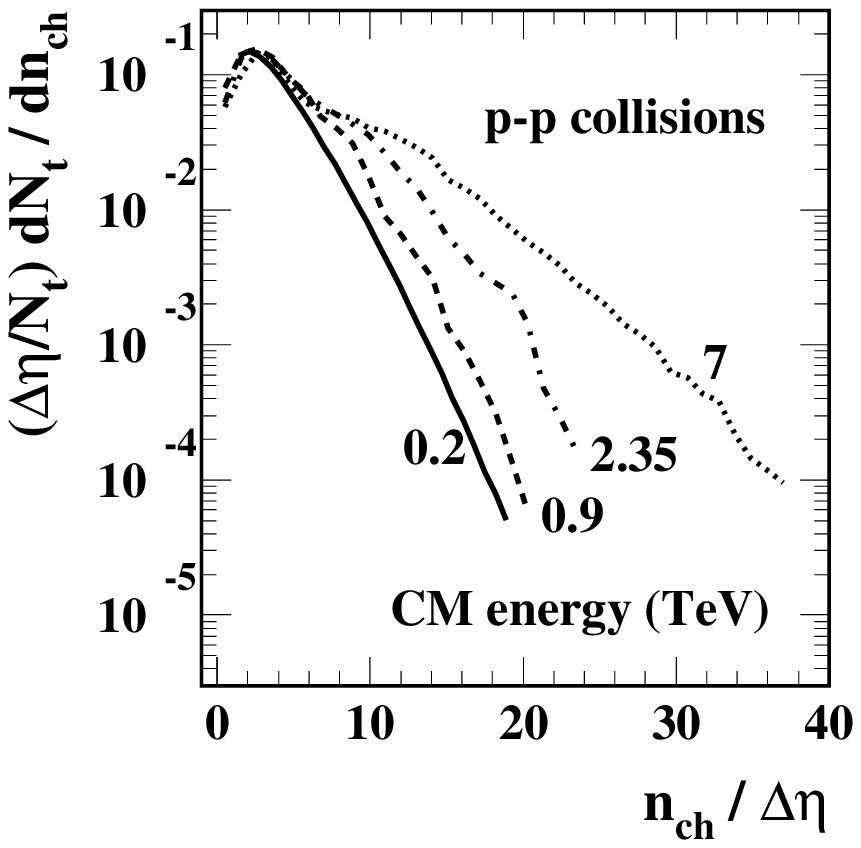}
\includegraphics[width=0.24\columnwidth,height=0.247\columnwidth]{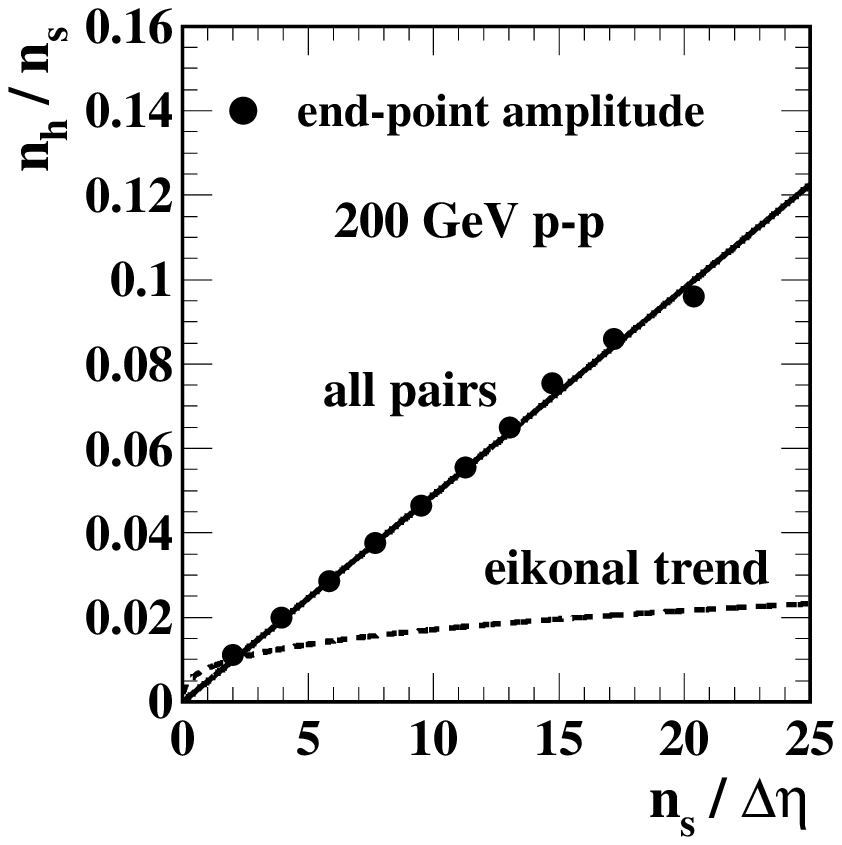}
\caption{
First: Single-particle spectra for seven $n_{ch}$ classes; 
Second: Scaled spectrum hard components $H(y_t,n_{ch})$; 
Third: Event distributions on $n_{ch}$ for several energies; 
Fourth: Hard-component multiplicity $n_h$ (dijet) trend on soft component $n_s$. $n_h$ is the integral (end-point amplitude) of measured $H(y_t,n_{ch})$ independent of shape. 
\label{ppspec}} 
\vspace{-.2in}
\end{figure}
Figure~\ref{ppspec} (first) shows rescaled $y_t$ spectra for seven multiplicity classes with $n_{ch} / \Delta \eta \approx 1.7,\ldots,19$. Fixed soft-component model $S_0$ is the asymptotic limit of spectra scaled by soft-component multiplicity $n_s$. Subtraction of $S_0$ and a second rescaling reveals hard components $H(y_t,n_{ch})$ scaled by $(n_s / \Delta \eta)^2$ (second panel) nearly independent of $n_{ch}$ approximated by fixed hard-component model $H_0(y_t)$. 
Soft-component multiplicity $n_s$ may serve as a proxy for {\em participant partons} (low-$x$ gluons) with substantial event-wise fluctuations (third panel). 
We observe (fourth panel) that $n_h \propto n_s^2$ (points), a trend inconsistent with that expected for the {\em eikonal model} (dashed curve $\propto n_s^{4/3}$) typically invoked in \pp\ Monte Carlo models~\cite{pythia,herwig}.
These 1D spectrum results provide the model functions and dijet systematics required to analyze and interpret the trigger-associated correlations presented below. 

\section{Systematics of minimum-bias p-p angular correlations}

Combinatoric minimum-bias (MB) angular correlations on angle differences $\eta_\Delta = \eta_1 - \eta_2$ and $\phi_\Delta = \phi_1 - \phi_2$ accepting all particle pairs (no $p_t$ cuts) can be described by a 2D model function including only a few elements~\cite{porter2,porter3,anomalous}. The principal correlation components are jet-related same-side (SS) 2D peak and away-side (AS) 1D peak on azimuth (back-to-back jets) and nonjet (NJ) quadrupole $\cos(2\phi_\Delta)$.
\begin{figure}[h]
\vspace{-.0in}
\centering
\includegraphics[width=0.24\columnwidth]{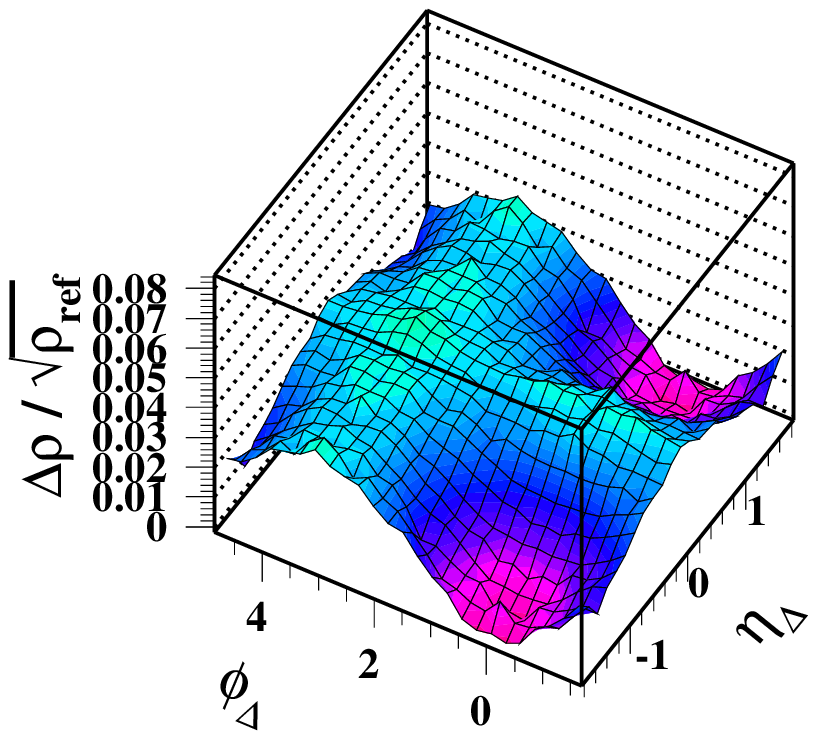}
\includegraphics[width=0.24\columnwidth]{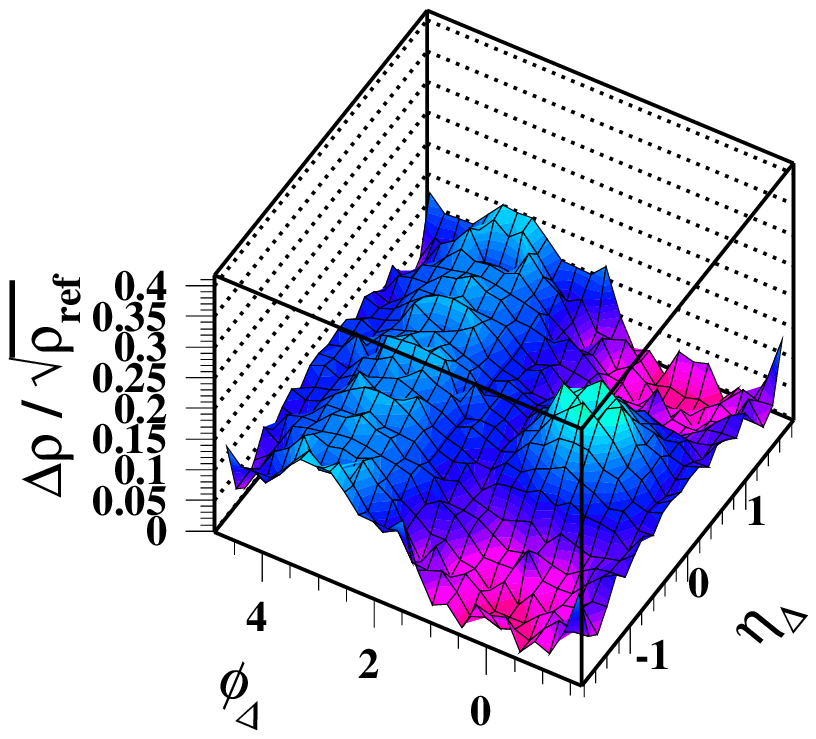}
\includegraphics[width=0.24\columnwidth,height=0.22\columnwidth]{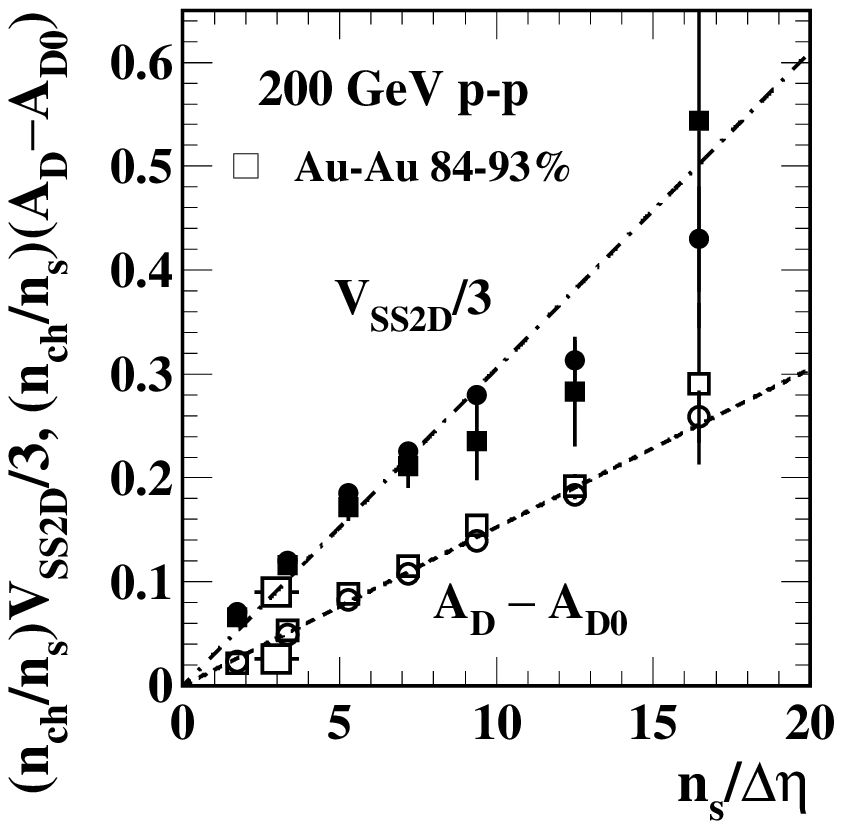}
\includegraphics[width=0.24\columnwidth,height=0.22\columnwidth]{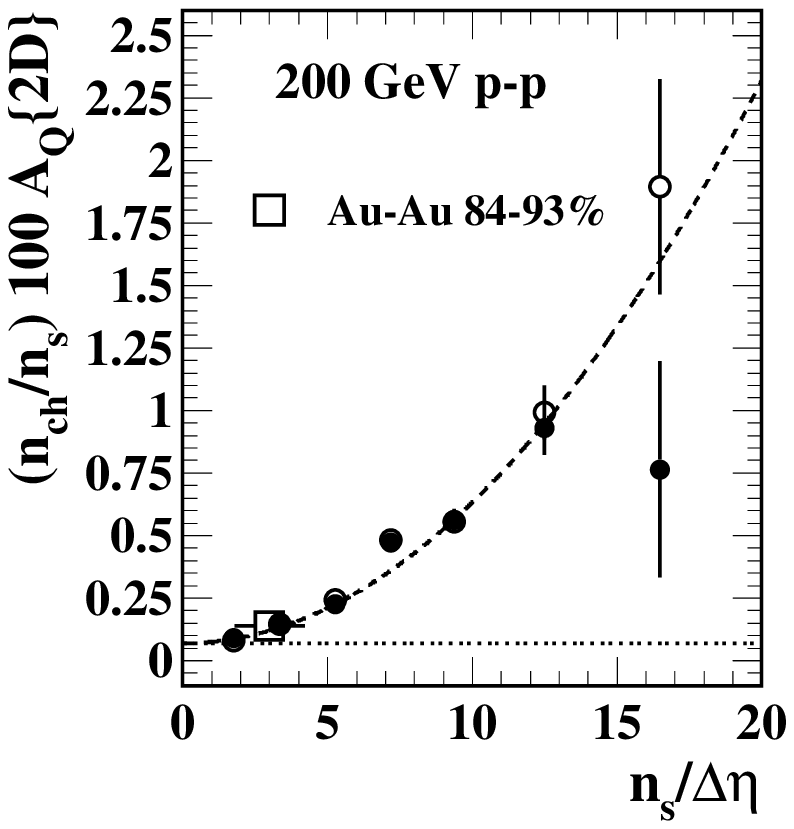}
\caption{
First, second: Jet-related and non-jet quadrupole angular correlations for multiplicity classes $n = 1$ and 6; Third:  Scaled amplitudes of jet-related structure vs soft multiplicity $n_s$; Fourth: Scaled nonjet quadrupole amplitude vs $n_s$
\label{angcorr}}  
\vspace{-.2in}
\end{figure}
Figure~\ref{angcorr} (first, second) shows angular correlations for multiplicity classes $n = 1$ and 6. Minor elements of the 2D model fits (proton fragment correlations, Bose-Einstein correlations, conversion electron pairs) have been subtracted leaving the jet-related components and the NJ quadrupole. The third panel shows trends on $n_s$ for jet-related amplitudes consistent with dijet number $n_j= 0.03 (n_s/2.5)^2$ (within $\Delta \eta = 2$) corresponding to pQCD dijet total cross section $\sigma_{dijet} = 2.5$ mb~\cite{fragevo}.
The \pp\ NJ quadrupole trend on $n_s$ can be predicted. The observed centrality trend for \auau\ collisions is $A_Q(b) \equiv \rho_0(b)v_2^2(b) \approx B\, N_{bin}(b) \epsilon_{optical}^2(b)$~\cite{davidhq}. For the non-eikonal \pp\ case $N_{bin} \rightarrow N_{part}^2$ and impact parameter $b$ is not an observable, so $n_{ch} A_Q(b)  \propto N_{part} N_{bin} \langle\epsilon_{optical}^2\rangle \propto N_{part}^3 \propto n_s^3$. Based on \pp\ dijet systematics we expect $(n_{ch}/n_s) A_Q(n_s) \propto n_s^2$, which is confirmed in the fourth panel.

\section{Trigger-associated (TA) two-component model (TCM)}

Based on \pp\ SP spectrum and 2D MB dijet angular correlation systematics we can construct a TCM for trigger-associated correlations~\cite{pptrig}. For each \pp\ collision {\em event type} (soft or hard) the hadron with the highest transverse rapidity $y_{tt}$ is the {\em trigger} particle. All other hadrons are {\em associated}, with rapidities $y_{ta}$. Definition of the TA TCM is an exercise in compound probabilities. The unit-normal 1D trigger spectrum for multiplicity class $n_{ch}$ denoted by $T(y_{tt},n_{ch}) \equiv [{1}/{N_{evt}(n_{ch})}] {dn_{trig}}/{y_{tt}dy_{tt}}$ is modeled by
\bea \label{trigspec}
T(y_{tt},n_{ch}) 
\hspace{-.0in }&=& \hspace{-.0in }  P_s(n_{ch})G_s(y_{tt})\, n_{ch} F_s(y_{tt}) 
\hspace{-.0in } +  \hspace{-.0in }    P_h(n_{ch})G_h(y_{tt})\, n_{ch} F_h(y_{tt}), 
\eea
where $P_x(n_{ch})$ is an {event-type} probability, $G_x(y_{tt})$ is a void (above $y_{tt}$) probability and $F_x(y_{tt})$ is a unit-normal SP spectrum for event-type $x = s$ (soft, no dijets) or $h$ (hard, at least one dijet), with $G_x(y_{tt})\, n_{ch} F_x(y_{tt}) \equiv T_x(y_{tt},n_{ch})$. The Poisson event-type probabilities are defined by $P_s = \exp(-n_j)$ and $P_h = 1 - P_s$. The void probabilities are defined by $G_x = \exp(- n_{x\Sigma})$, where $n_{x\Sigma}$ is the appropriate spectrum integral above $y_{tt}$.
\begin{figure}[h]
\centering
\includegraphics[width=0.24\columnwidth,height=0.22\columnwidth]{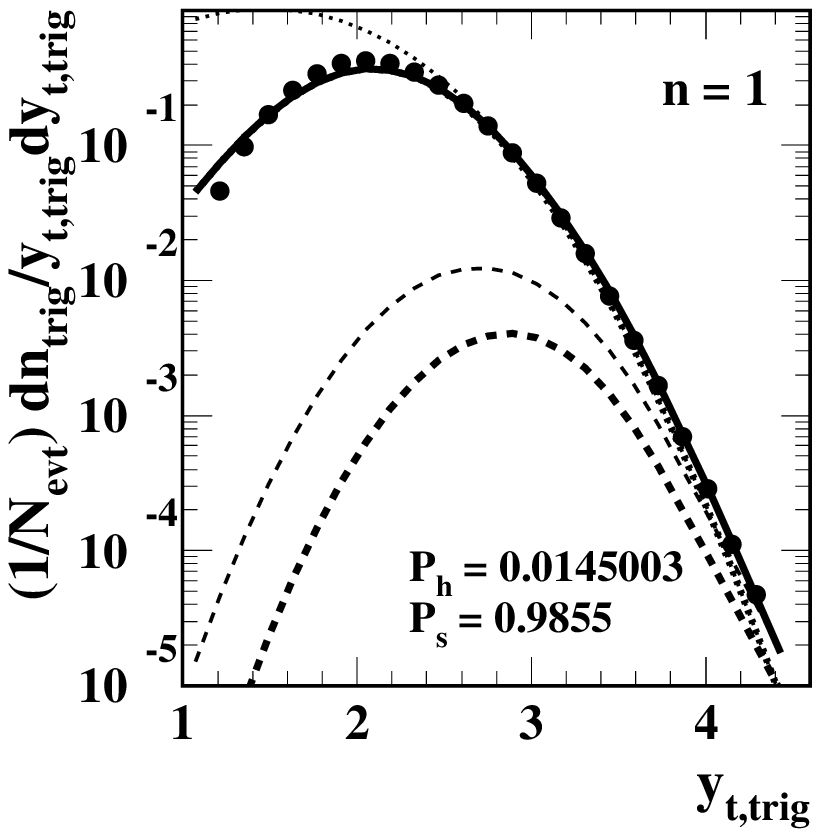}
\includegraphics[width=0.24\columnwidth,height=0.22\columnwidth]{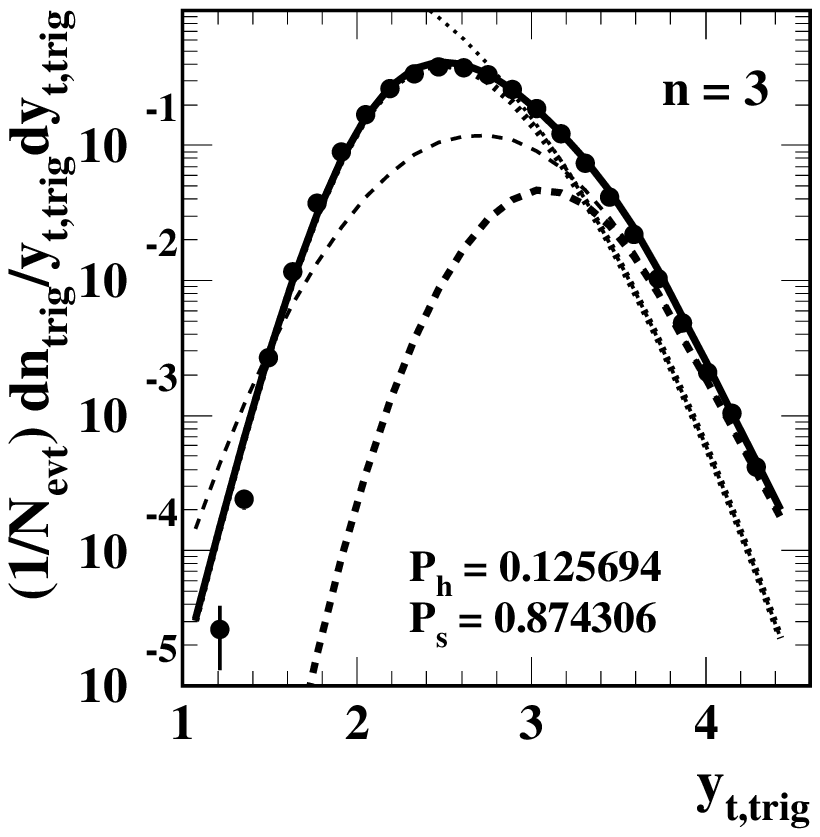}
\includegraphics[width=0.24\columnwidth,height=0.22\columnwidth]{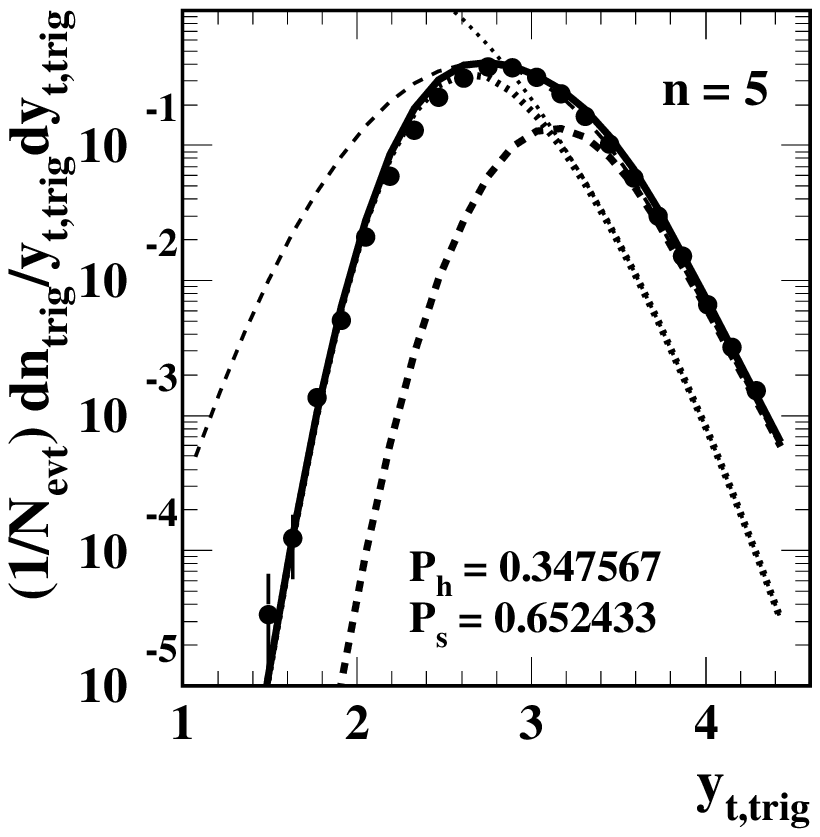}
\includegraphics[width=0.24\columnwidth,height=0.22\columnwidth]{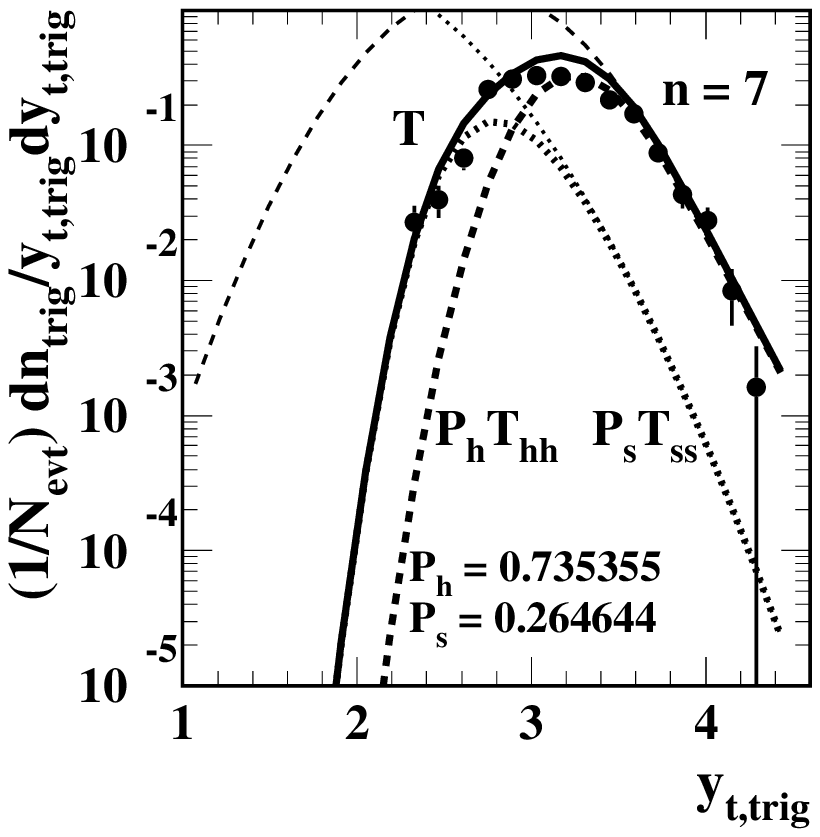}
\caption{
Trigger spectrum data (points) and TCM (curves) for n = 1, 3, 5, 7.
\label{tatcm}} 
\vspace{-.2in}
\end{figure}
Figure~\ref{tatcm} shows trigger spectra (points) for four multiplicity classes. Solid curves $T(y_{tt},n_{ch})$ are defined by Eq.~(\ref{trigspec}). The other curves refer to TCM trigger-spectrum components. The TCM describes the trigger spectra well.

The unit-normal 2D TA distribution for event-type $x$ and multiplicity class $n_{ch}$ is {\em joint probability} $F_x(y_{ta},y_{tt}) \equiv T_x(y_{tt}) A_x(y_{ta}|y_{tt})$, where the {\em chain rule} for compound probabilities has been invoked. $A_x(y_{ta}|y_{tt})$ is the {\em conditional probability} that an associated particle is emitted at $y_{ta}$ in an event of type $x$ given a trigger at $y_{tt}$ with probability $T_x(y_{tt})$. The TA TCM is then
\bea \label{tadist}
F(y_{ta},y_{tt}, n_{ch})
&=&    P_s (n_{ch}) T_s(y_{tt})A_s(y_{ta}|y_{tt}) 
+ P_h (n_{ch}) T_h(y_{tt}) A_h(y_{ta}|y_{tt}),~~
\eea
where the TCM $A_x$ are formed from the SP-spectrum TCM elements with certain {marginal constraints}~\cite{pptrig}. Hard component $H_h(y_{ta}|y_{tt})$ of  $A_h(y_{ta}|y_{tt})$ represents the sought-after momentum correlation structure of MB jets.

\section{Measured trigger-associated correlations}

Trigger-associated correlations can be presented both as joint probabilities $F(y_{ta},y_{tt},n_{ch})$ and as conditional probabilities $A(y_{ta}|y_{tt},n_{ch}) = F / T$ using the chain rule for joint probabilities.
\begin{figure}[h]
\centering
\includegraphics[width=0.24\columnwidth]{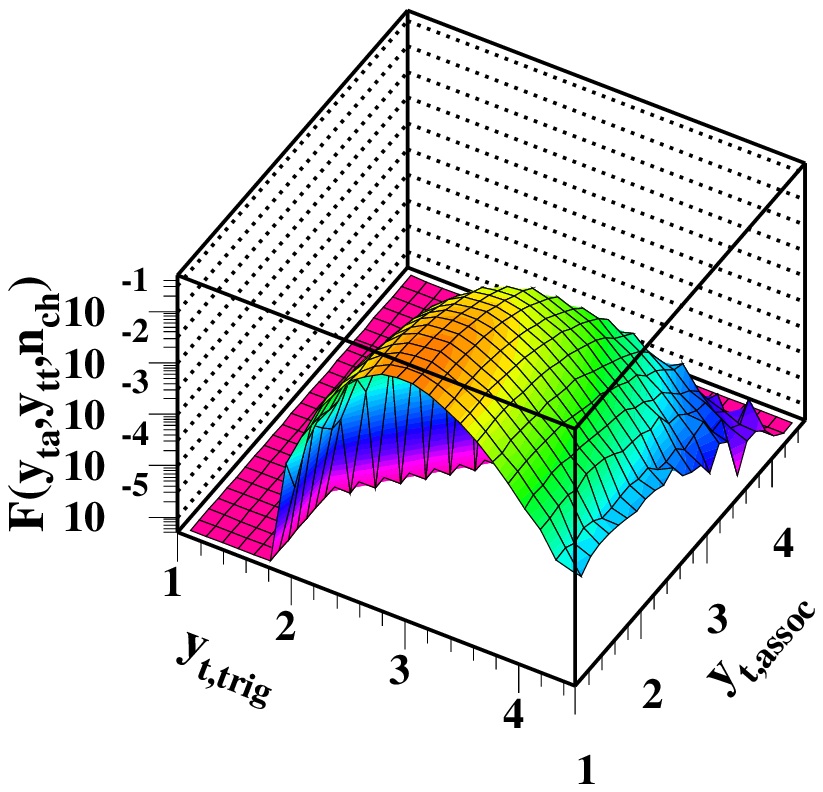}
\includegraphics[width=0.24\columnwidth]{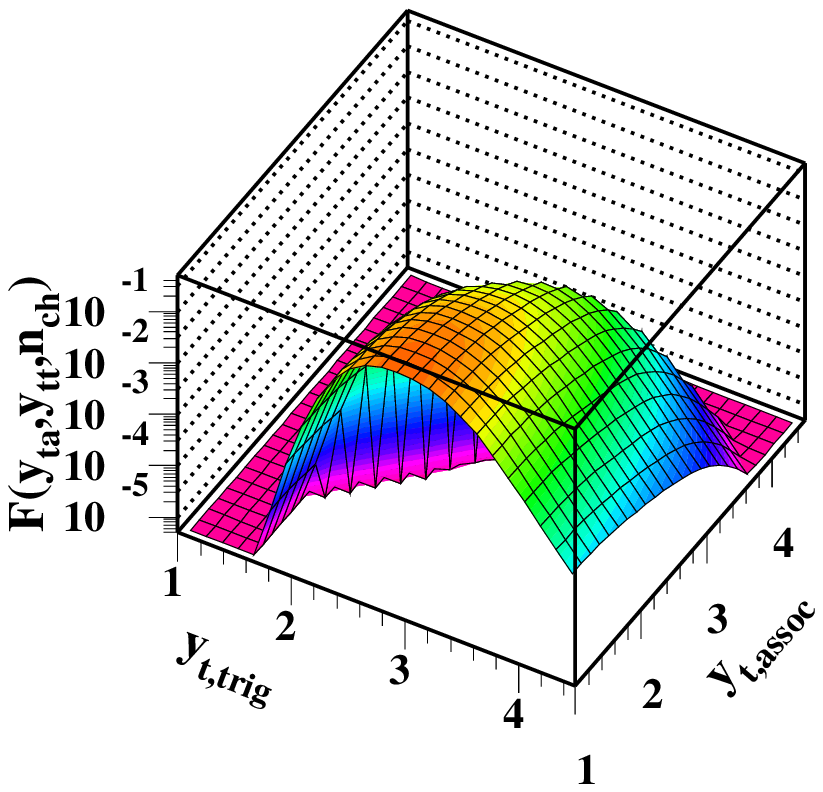}
\includegraphics[width=0.24\columnwidth,height=0.23\columnwidth]{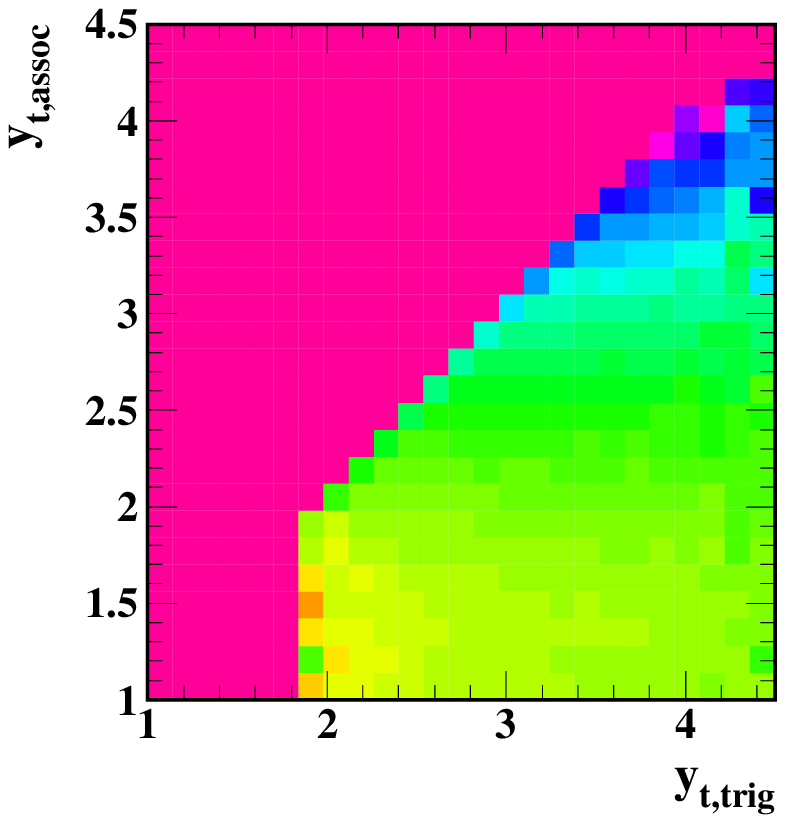}
\includegraphics[width=0.24\columnwidth,height=0.23\columnwidth]{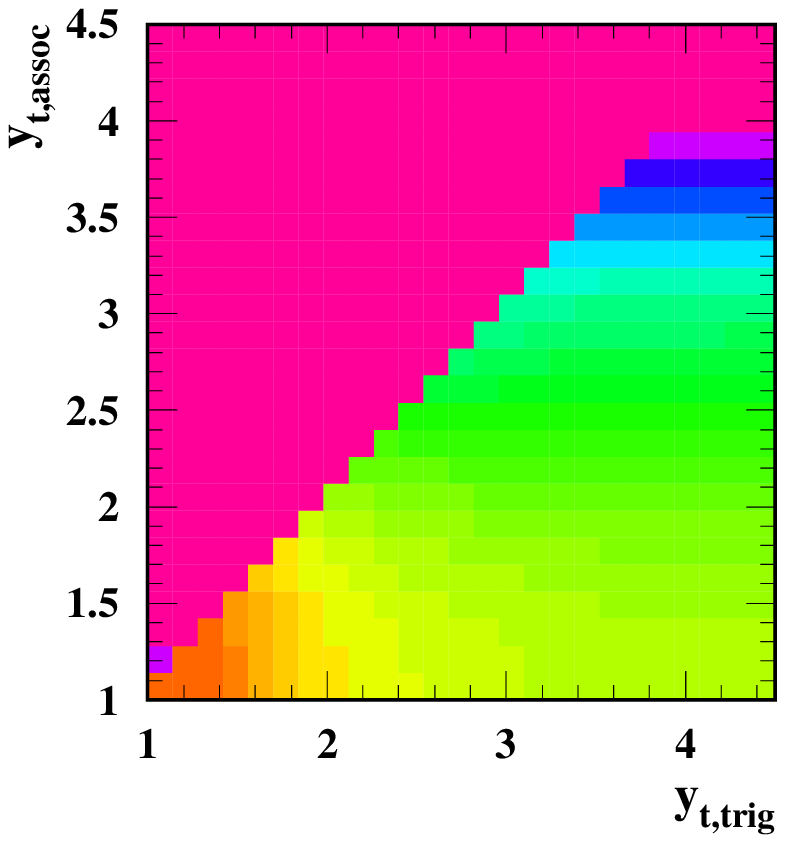}
\caption{
Left: TA correlations F for multiplicity class $n = 6$ and for data and TCM (first and second respectively);
Right: Same for conditional correlations $A = F/T$.
\label{tadata}} 
\vspace{-.2in}
\end{figure}
Figure~\ref{tadata} (left) shows the measured joint distribution $F$ for $n = 6$ (first) and its corresponding TCM (second). Figure~\ref{tadata} (right) shows the measured conditional distribution $A$ (third) and its TCM (fourth). In both cases the agreement is good below $y_{ta} \approx 2.5$. TCM hard component $H_0'$ is based on a simple factorization approximation and plays no role in extraction of the data hard components described below.  The jet-related hard component dominates TA structure for $y_{ta}$, $y_{tt} > 2.5$. The data and TCM hard components differ substantially.

\section{Extracting the TA hard component}

Dividing Eq.~(\ref{tadist}) by Eq.~(\ref{trigspec}) we obtain the total conditional distribution
\bea \label{tacond}
A(y_{ta}|y_{tt}, n_{ch})
&=&    R_s (y_{tt},n_{ch}) A_s(y_{ta}|y_{tt}) 
+ R_h (y_{tt},n_{ch}) A_h(y_{ta}|y_{tt}),~~
\eea
where the $R_x \leq 1$ are {\em trigger fractions}. The TCM conditional distributions are $A_s = S_0''$ and $A_h = p_s' S_0' + p_h' H_0'$ for $y_{ta} < y_{tt}$, with primes on $X_0'$ denoting the effects of marginal constraints as described in Ref.~\cite{pptrig}, and $p_x' = n_x' / (n_{ch} - 1)$.
Given that expression we can isolate the hard component of the TA conditional distribution by subtracting the TCM soft components
\bea \label{hardcomp}
H_h'(y_{ta}|y_{tt},n_{ch}) \hspace{-.02in} &=& \hspace{-.02in} \frac{n_{ch}-1}{R_h} [A(y_{ta}|y_{tt}) - R_s S''_0(y_{ta}|y_{tt}) -  R_h p'_s S'_0(y_{ta}|y_{tt})],~~~
\eea
the hard component (dijet momentum structure) per hard event. All subtractions use the same soft-component models derived from SP spectra.

\begin{figure}[h]
\centering
\includegraphics[width=0.24\columnwidth]{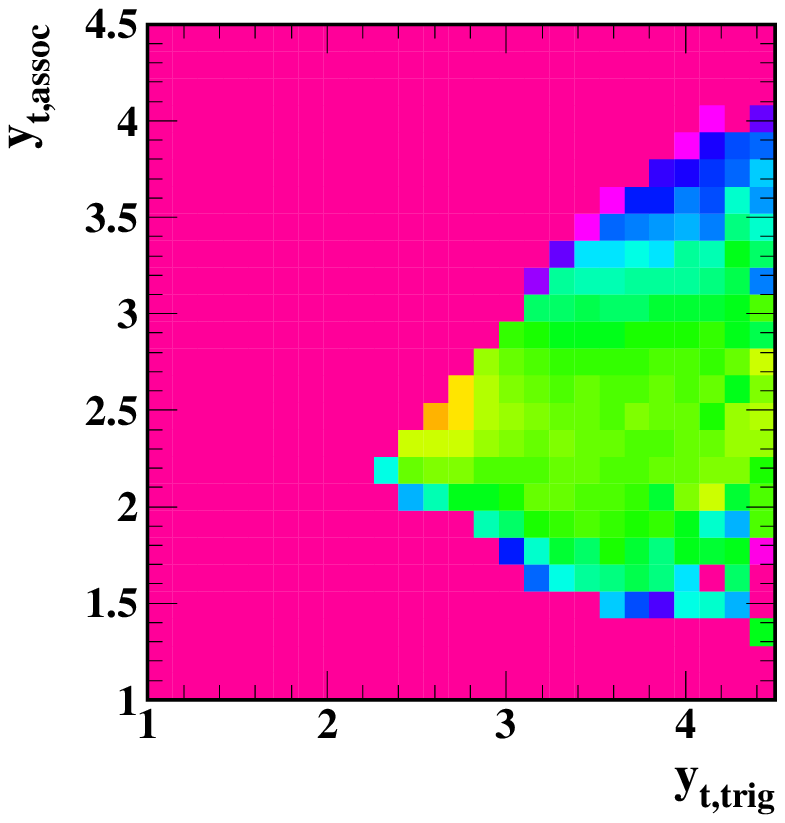}
\includegraphics[width=0.24\columnwidth]{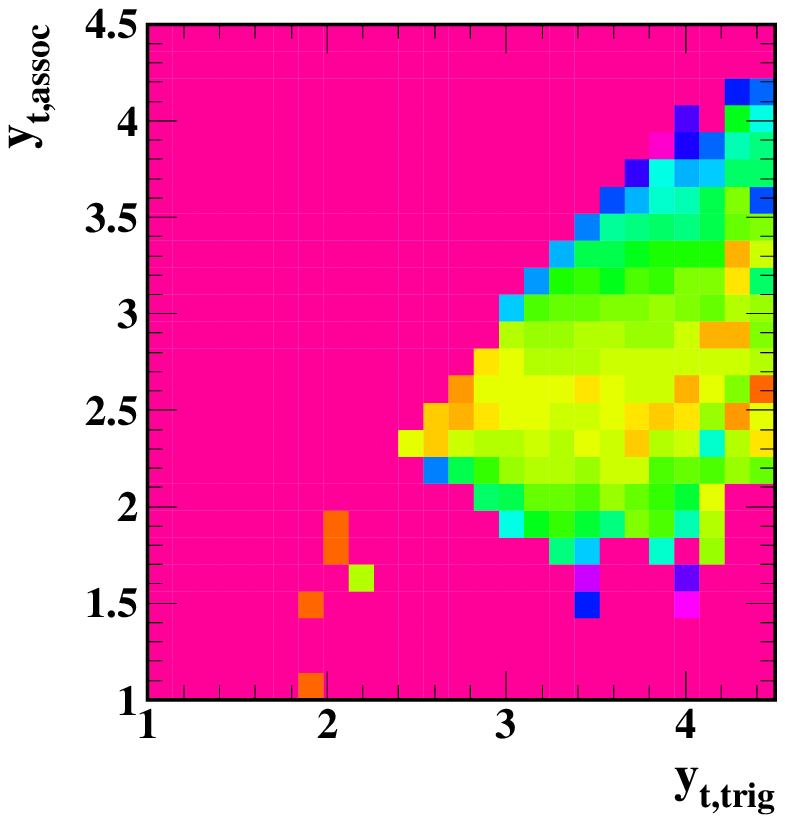}
\includegraphics[width=0.24\columnwidth]{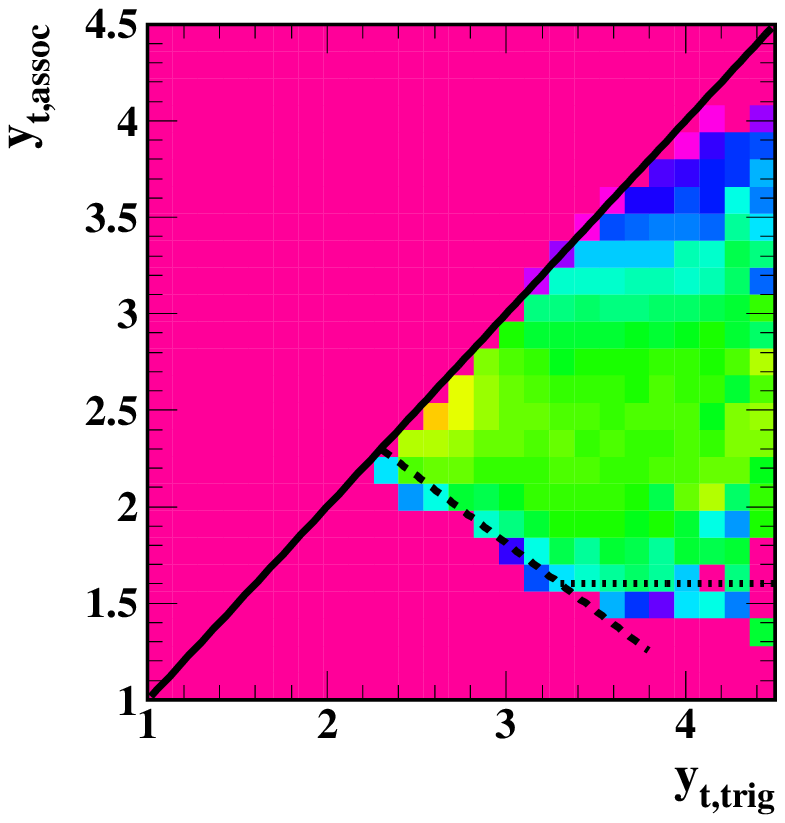}
\includegraphics[width=0.24\columnwidth]{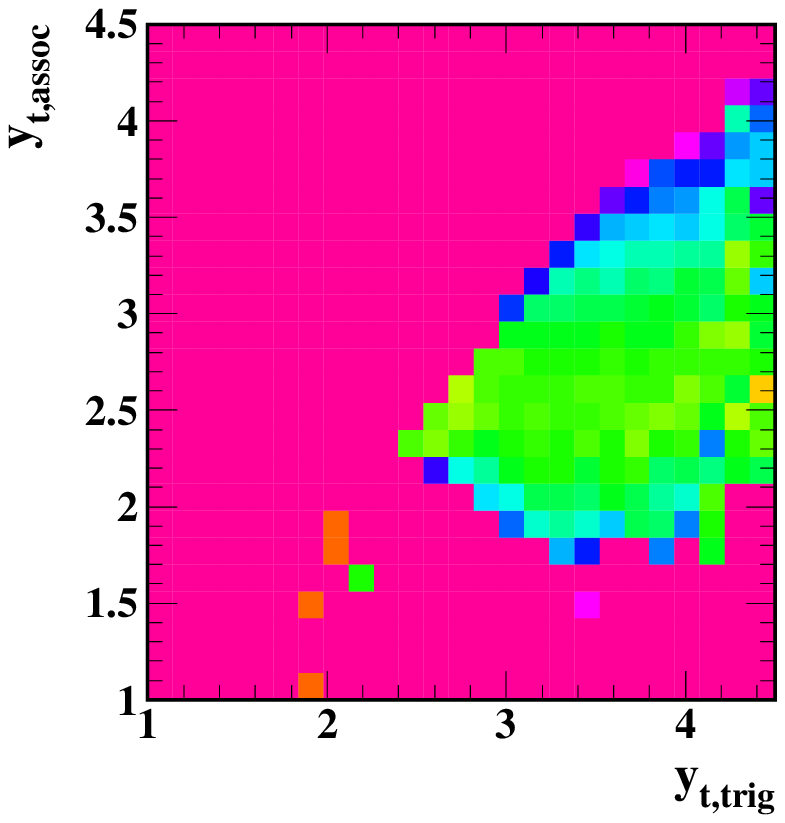}
\caption{
Left: Per-hard-event hard component $H_h'(y_{ta}|y_{tt},n_{ch})$ for multiplicity classes $n = 2$ and 6;
Right: The same data scaled by number of dijets per hard event $n_j / P_h$ to yield the per-dijet hard component. Lines are discussed in the text.
\label{hardcomp}} 
\vspace{-.2in}
\end{figure}

Figure~\ref{hardcomp} (left) shows hard components $H_h'(y_{ta}|y_{tt})$ for multiplicity classes $n = 2$, 6 (first and second respectively). The jet structure {\em per hard event} increases substantially with $n_{ch}$ because the probability that one or more additional dijets accompanies a triggered dijet (multiple parton interactions or MPI) becomes substantial. We can divide the left panels by the number of dijets per hard event $n_j / P_h$ to obtain the right panels. The resulting {\em per-dijet} structure appears to be approximately independent of $n_{ch}$ (universal). Universality is consistent with the $n_j(n_s)$ trend inferred from SP spectra.

\section{TA azimuth dependence and the transverse region or TR}

The azimuth structure of TA correlations relative to the trigger direction is of interest for several reasons including ``underlying event'' (UE) studies.
\begin{figure}[h]
\centering
\includegraphics[width=0.24\columnwidth]{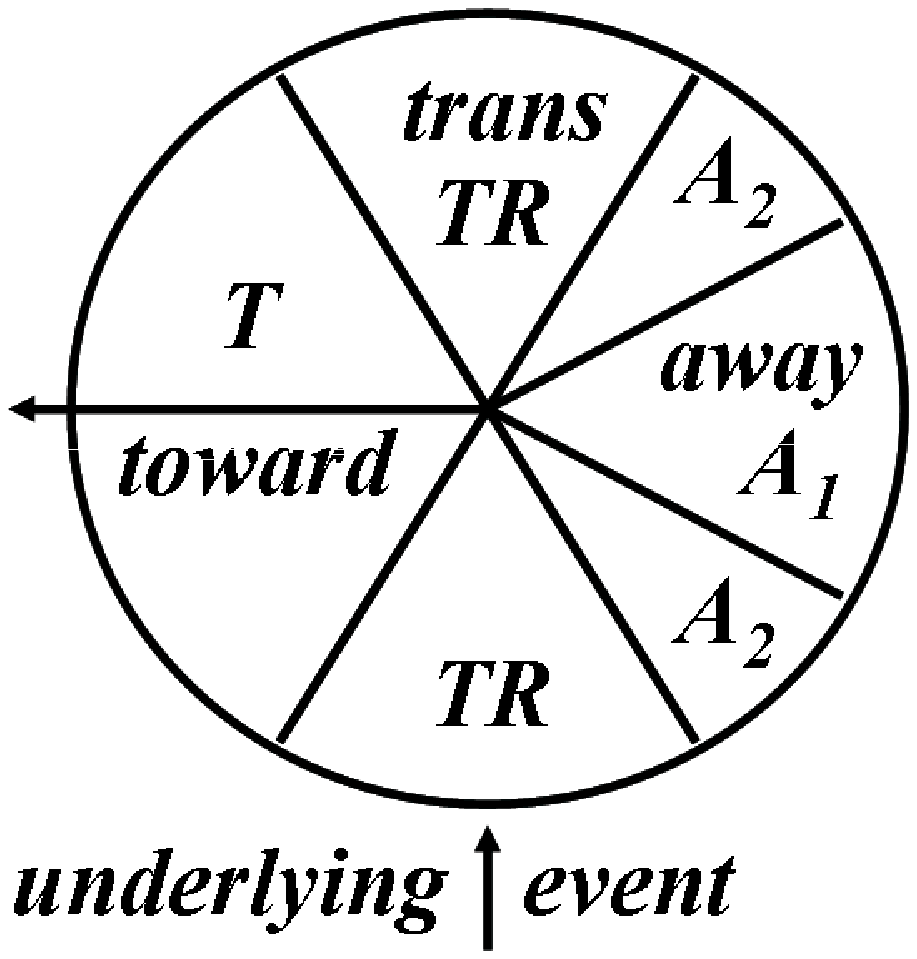}
\includegraphics[width=0.24\columnwidth]{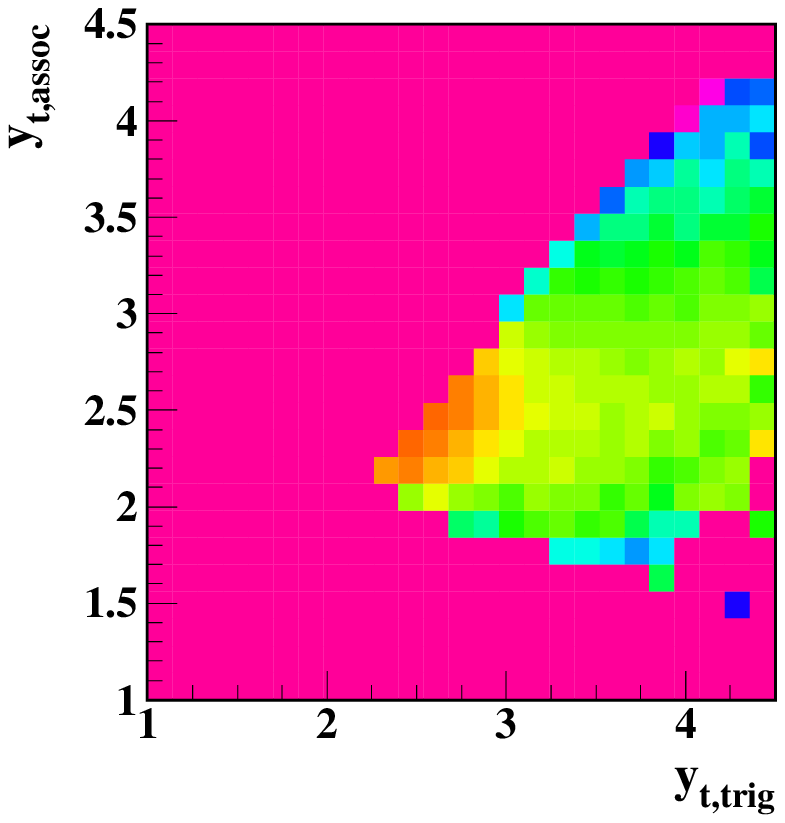}
\includegraphics[width=0.24\columnwidth]{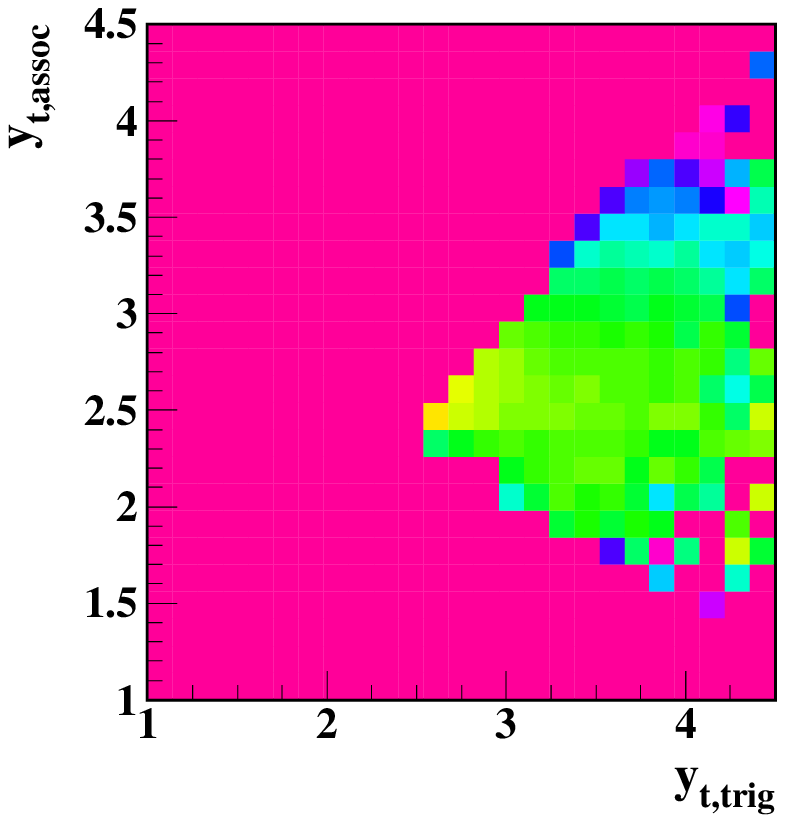}
\includegraphics[width=0.24\columnwidth]{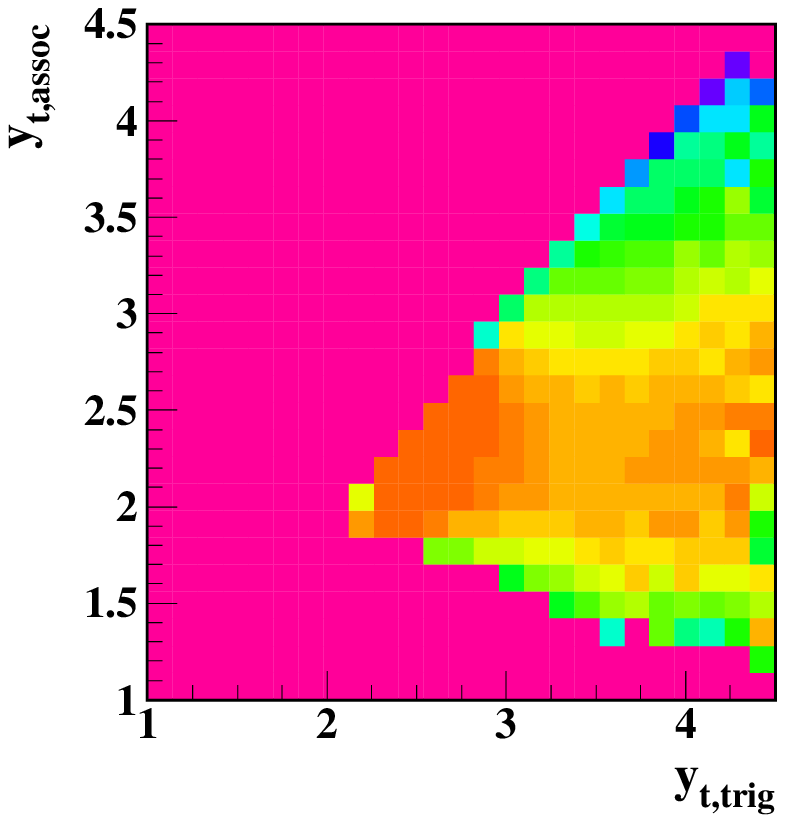}
\caption{ 
First: Conventional definition of azimuth regions for underlying event analysis;
Second, Third, Fourth: TA hard components for toward T, transverse TR and away A averaged over multiplicity classes $n = 2$, 3, 4 to minimize MPI.
\label{azimuth}} 
\vspace{-.2in}
\end{figure}
Figure~\ref{azimuth} (first) shows the conventional azimuth partition relative to trigger direction (arrow) into three equal regions: ``toward'' (T), ``transverse'' (TR) and ``away'' (A). In some studies the A region is split into two parts $A_1$ and $A_2$ as shown. In conventional UE analysis it is assumed that the triggered dijet does not contribute to the TR, which region should therefore permit unbiased access to the UE {\em complementary to the triggered dijet}~\cite{cdfue,rick}.

Figure~\ref{azimuth} (second, third, fourth) shows TA hard components per hard event for T, TR and A regions respectively, averaged over lower multiplicity classes $n = 2$, 3, 4 to reduce dijet pileup (MPI) to less than 15\%. Those data averaged over azimuth are equivalent to Fig.~\ref{hardcomp} (right). Most notable is the substantial triggered-dijet contribution to the TR region (third panel), contradicting a common  UE assumption. Compared to the T region (second) the A region (fourth) is both significantly softer {\em and} harder. The A region must be harder on average to compensate the trigger particle excluded from conditional distribution $A$ in the T region. The A region is also softer on average because of trigger bias to lower-energy jets due to initial-state $k_t$ effects and toward a softer fragmentation cascade within those jets.

\section{The underlying event and multiple parton interactions}

\begin{figure}[h]
\centering
\includegraphics[width=0.24\columnwidth,height=0.23\columnwidth]{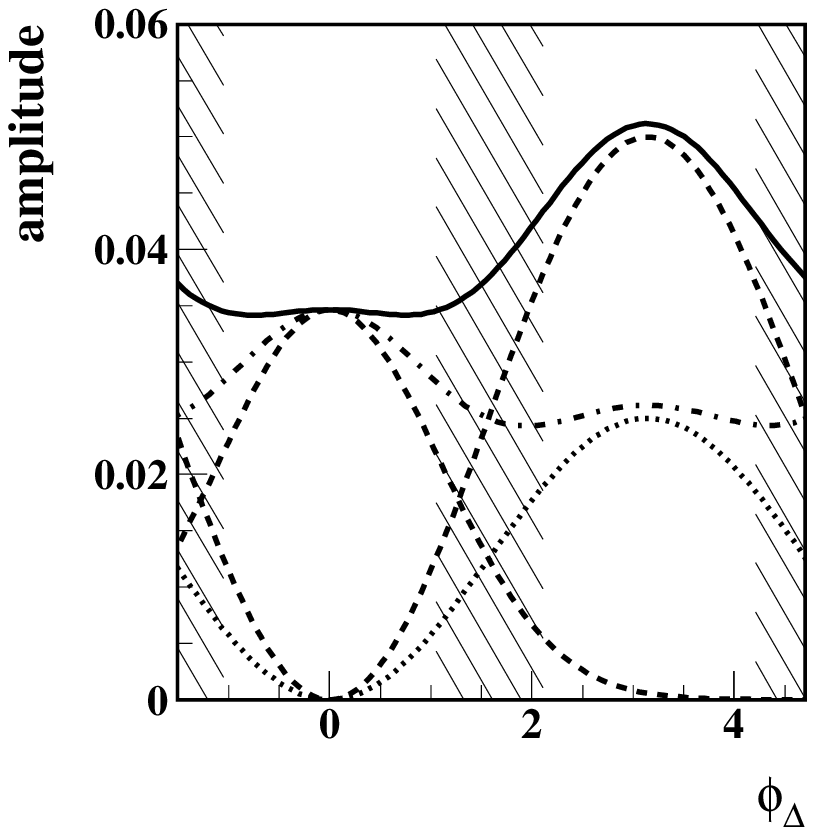}
\includegraphics[width=0.24\columnwidth,height=0.23\columnwidth]{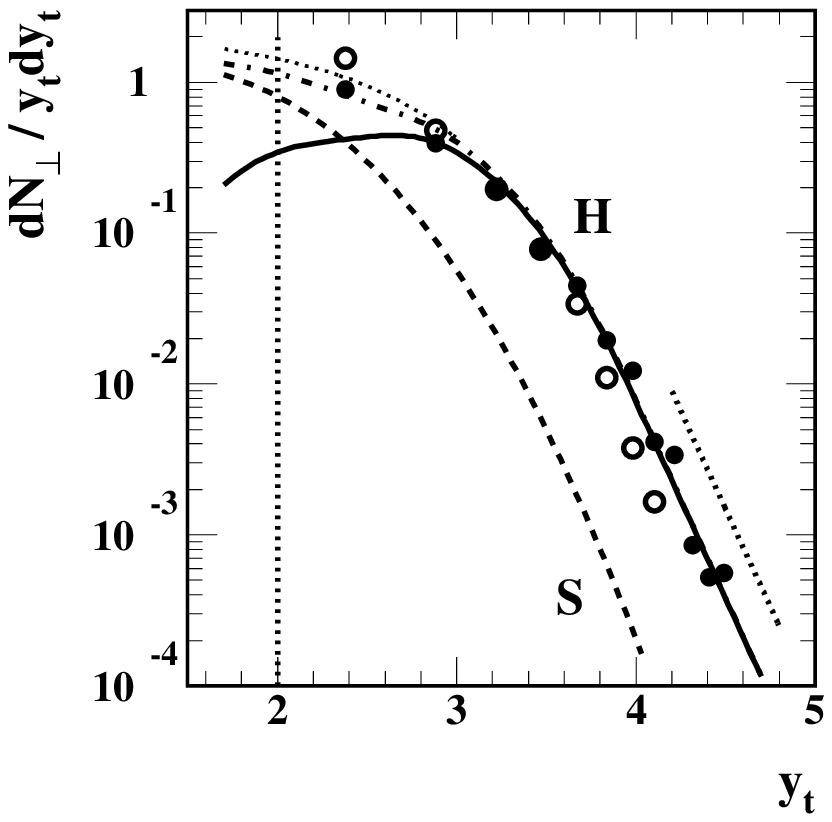}
\includegraphics[width=0.24\columnwidth,height=0.233\columnwidth]{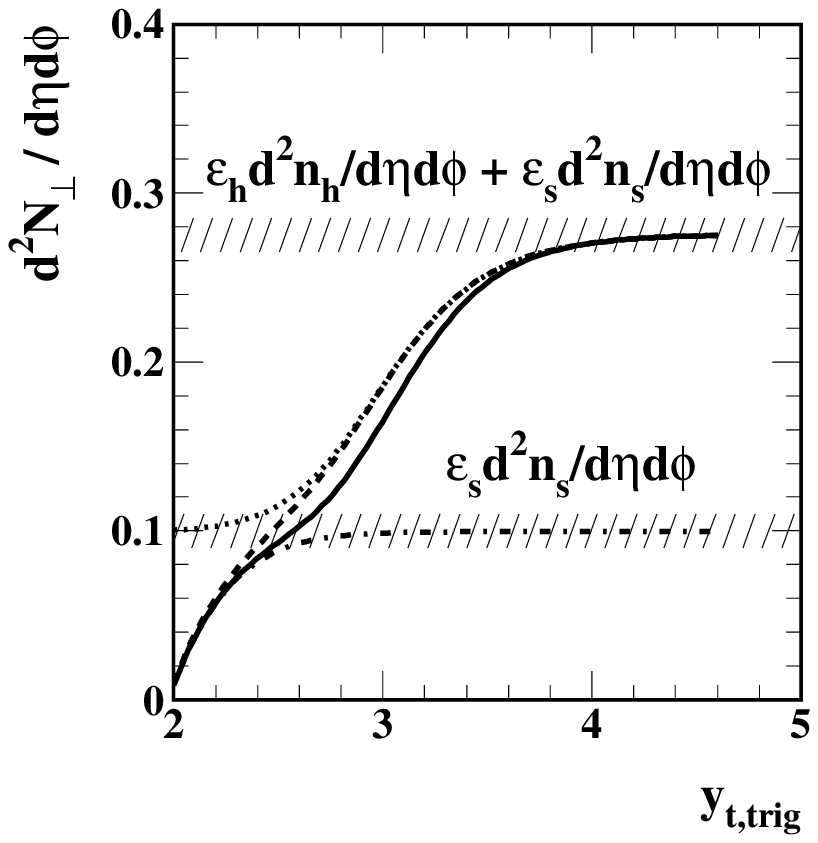}
\includegraphics[width=0.24\columnwidth,height=0.233\columnwidth]{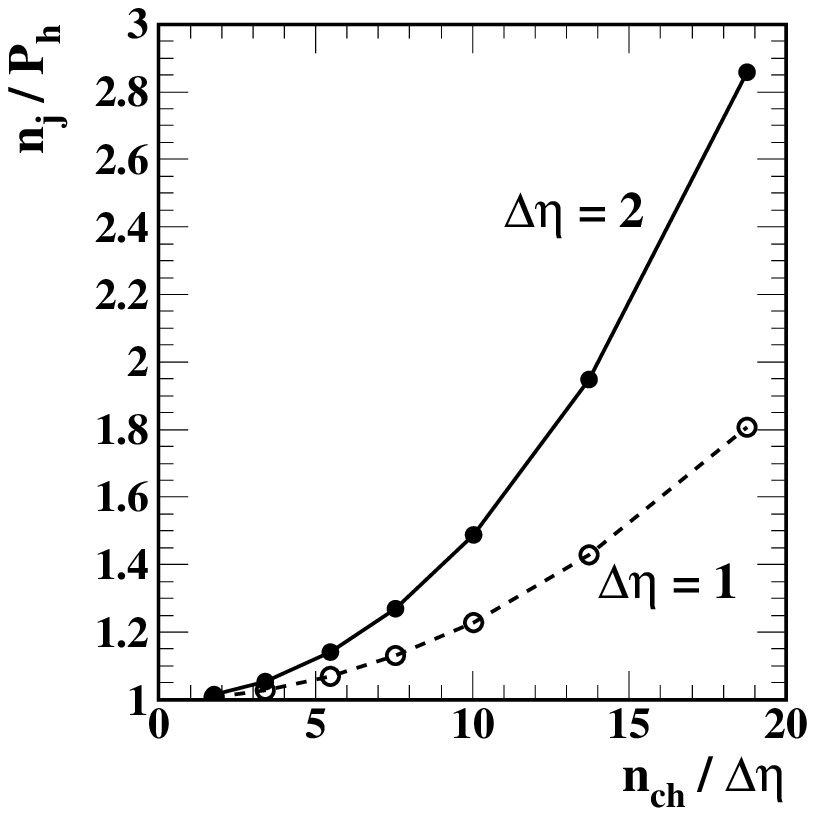}
\caption{
First: Model fit to 2D MB jet angular correlations (curves) projected onto 1D azimuth showing substantial jet contribution to the TR (hatched);
Second: Spectrum of $N_\perp$ yield in the TR (points,~\cite{cdfue}) showing jet-related hard component (curve $H$);
Third: Simulated $N_\perp$ density vs jet trigger condition showing increase to saturation due to selection of low-multiplicity hard events~\cite{pptheory};
Fourth: Number of jets per hard event $n_j / P_h$ vs $n_{ch}$ inferred from SP spectrum systematics~\cite{ppprd,pptheory}.
\label{uesys}} 
\vspace{-.2in}
\end{figure}

Other issues emerge for conventional UE analysis. Based on MB dijet angular correlations as in Fig.~\ref{angcorr} (left) we expect a substantial contribution to the TR from any dijet~\cite{pptheory}. Figure~\ref{uesys} (first) shows a projection onto azimuth of the model fit to Fig.~\ref{angcorr} (first) approximating MB jet structure from non-single-diffractive \pp\ collisions. There is a substantial overlap of SS and AS jet peaks and resulting strong jet contribution to the TR. Figure~\ref{uesys} (second) shows $N_\perp$ spectra from the TR described by the TCM of Eq.~(\ref{ppspec}) with the amplitude of (jet) hard-component $H$ as expected for hard (triggered) events.
Fig.~\ref{angcorr} (third) shows the TR $N_\perp$ density vs trigger condition $y_{t,trig}$. The increase to a saturation value is conventionally attributed to MPI. However, a study based on the TCM for SP spectra reveals that the $N_\perp$ increase results from a dijet contribution to the TR for hard events with low ($\approx$ NSD) multiplicities where the incidence of MPI is negligible~\cite{pptheory}. Fig.~\ref{angcorr} (fourth) shows the calculated dijet number per hard event vs multiplicity. For NSD \pp\ collisions ($n_{ch} / \Delta \eta \approx 2.5$) the MPI rate is only a few percent.
From TA and angular-correlation analysis we conclude that application of a trigger $y_{t,trig}$ (jet) condition in UE analysis selects for jets within mainly low-multiplicity ($\approx$ NSD) hard events. Applying an $n_{ch}$ condition instead would select for multiple MB dijets (MPI) in higher-multiplicity events.

\section{Kinematic limits on physical MB jet fragment production}

The results in Figs.~\ref{hardcomp} (lines in third) and \ref{azimuth} (second and fourth panels) reveal the kinematic limits of minimum-bias jet fragment production: Trigger hadrons extend down to $\approx 1$ GeV/c ($y_{tt} \approx 2.7$), and associated hadrons extend down to $\approx 0.35$ GeV/c (AS, $y_{ta} \approx 1.5$) or 0.5 GeV/c (SS, $y_{ta} \approx 2$). We conjecture that this also represents the low-hadron-momentum (and large-angle) structure of high-parton-energy jets, the common base of any dijet. Higher-energy jets contain a few additional high-momentum hadrons located close to the dijet axis and therefore outside the TR. TA correlation analysis could be extended to \aa\ collisions to verify the strong contribution from MB jets (minijets) even in more-central \auau\ collisions~\cite{anomalous,fragevo,jetspec}.

These TA results are consistent with measured FFs from LEP,
HERA and CDF and with a pQCD parton spectrum
that predicts measured dijet production~\cite{eeprd,pptheory} and the shape of the MB spectrum hard component~\cite{fragevo}. The MB-jet-related SS 2D peak volume is also consistent with pQCD predictions~\cite{jetspec}.
Conventional trigger-associated $p_t$ {\em cuts} invoked in \aa\ dihadron correlation analysis accept only a small fraction of the actual dijet number and jet fragments and, combined with so-called ZYAM subtraction of a combinatorial background, produce an unphysical picture of dijets in nuclear collisions in which jet structure is minimized and distorted~\cite{tzyam}.

\section{Summary}

The two-component (soft + hard) model (TCM) of hadron production in high-energy nuclear collisions works remarkably well. Based on various comparisons with theory the soft component represents fragments from projectile nucleons (their gluon constituents), and the hard component represents dijet fragments from large-angle-scattered partons (gluons).

In this study the TCM has been applied to MB trigger-associated (TA) correlations for several charge multiplicity classes of 200 GeV \pp\ collisions. A conditional hard component $H_h(y_{ta}|y_{tt})$ has been extracted by analogy with TCM analysis of single-particle spectra. The TA hard component reveals the kinematic limits of jet fragment production and is directly comparable with measured jet fragmentation functions from \ee\ collisions.

These TA correlation results have implications for underlying-event (UE) analysis. Consistent with MB angular-correlation analysis the TA results confirm that the triggered dijets make a strong contribution to the transverse region or TR, contradicting conventional UE assumptions. The increase of the $N_\perp$ charge multiplicity in the TR with jet trigger $y_{t,trig}$ results not from multiple parton interactions (MPI) but from increased probability of low-multiplicity hard events including only a single dijet. The MPI rate is increased by selecting instead higher event multiplicities $n_{ch}$. The physical UE is then the MPI rate determined by $n_{ch}$ plus the TCM soft component.

\vskip .1in
This work was supported in part by the Office of Science of the U.S.\ DOE under grant DE-FG03-97ER41020. 



\begin{thebibliography}{99}

\bibitem{ppprd} J.~Adams {\it et al.}  (STAR Collaboration),
  Phys.\ Rev.\  D {\bf 74}, 032006 (2006).

\bibitem{pythia}  T. Sj\"ostrand and M. van Zijl, Phys. Rev. D {\bf 36}, 2019 (1987);
T.~Sj\"ostrand,
Comput.\ Phys.\ Commun.\  {\bf 82}, 74 (1994); 
T.~Sj\"ostrand, L.~L\"onnblad, S.~Mrenna and P.~Skands,
hep-ph/0308153.

\bibitem{herwig}  M.~Bahr, S.~Gieseke, M.~A.~Gigg, D.~Grellscheid, K.~Hamilton, O.~Latunde-Dada, S.~Platzer and P.~Richardson {\it et al.},
  Eur.\ Phys.\ J.\ C {\bf 58}, 639 (2008).

\bibitem{porter2} R.~J.~Porter and T.~A.~Trainor  (STAR Collaboration),
  J.\ Phys.\ Conf.\ Ser.\  {\bf 27}, 98 (2005).

\bibitem{porter3}  R.~J.~Porter and T.~A.~Trainor  (STAR Collaboration),
  PoS C {\bf FRNC2006}, 004 (2006).

\bibitem{anomalous}  G.\ Agakishiev, {\it et al.} (STAR Collaboration),
  Phys.\ Rev.\ C {\bf 86}, 064902 (2012).

\bibitem{fragevo}    T.~A.~Trainor,
  Phys.\ Rev.\  C {\bf 80}, 044901 (2009).

\bibitem{davidhq}  D.~T.~Kettler  (STAR collaboration),
  Eur.\ Phys.\ J.\  C {\bf 62}, 175 (2009).

\bibitem{pptrig} T.~A.~Trainor and D.~J.~Prindle,
  arXiv:1307.1819.

\bibitem{cdfue}   T.~Affolder {\it et al.}  (CDF Collaboration),
  Phys.\ Rev.\ D {\bf 65}, 092002 (2002).

\bibitem{rick} R.~Field,
  Acta Phys.\ Polon.\ B {\bf 42}, 2631 (2011).

\bibitem{eeprd}   T.~A.~Trainor and D.~T.~Kettler,
  Phys.\ Rev.\ D {\bf 74}, 034012 (2006).

\bibitem{pptheory}  T.~A.~Trainor,
Phys.\ Rev.\ D {\bf 87}, 054005 (2013).

\bibitem{jetspec}   T.~A.~Trainor and D.~T.~Kettler,
  Phys.\ Rev.\ C {\bf 83}, 034903 (2011).

\bibitem{tzyam} T.~A.~Trainor,
  Phys.\ Rev.\  C {\bf 81}, 014905 (2010).


\end{thebibliography}
\end{document}